# Observations of super-resolution using phase-controlled coherent photons in a delayed-choice quantum eraser scheme

Sangbae Kim and Byoung S. Ham
School of Electrical Engineering and Computer Science, Gwangju Institute of Science and Technology,
123 Chumdangwagi-ro, Buk-gu, Gwangju 61005, South Korea
(December 6, 2023; bham@gist.ac.kr)

**Abstract**
Super-resolution overcoming the standard quantum limit has been intensively studied for quantum sensing applications of precision target detection over the last decades. Not only higher-order entangled photons but also phase-controlled coherent photons have been used to demonstrate the super-resolution. Due to the extreme inefficiency of higher-order entangled photon-pair generation and ultralow signal-to-noise ratio, however, quantum sensing has been severely limited. Here, we report observations of coherently excited super-resolution using phase-controlled coherent photons in a delayed-choice quantum eraser scheme. Using phase manipulations of the quantum erasers, super-resolution has been observed for higher-order intensity correlations between them, satisfying the Heisenberg limit in phase resolution. This new type of precision phase-detection technique opens the door to practical applications of quantum sensing compatible with current technologies based on coherence optics.

**Introduction**
In classical physics, the Rayleigh criterion defines the phase resolution of coherent light in an interferometric system for the first-order intensity correlation, resulting in the so-called diffraction limit [1]. Higher-order intensity correlations enhance the resolution further according to the standard quantum limit (SQL) [2-5]. On the other hand, the Heisenberg limit in quantum sensing enhances the phase resolution beyond SQL when the probe photons are entangled [3-14] or squeezed [15-17]. Over the last several decades, quantum sensing overcoming SQL has been investigated for super-resolution as well as super-sensitivity using photonic de Broglie waves (PBWs) [6-8], higher-order entangled photon pairs, i.e., N00N states [9-14], squeezed light [15-17], and orbital angular momenta [18]. Regarding super-resolution, not only nonclassical light but also coherent light has been demonstrated for overcoming SQL, where the phase control of coherent photons plays a key role in the super-resolution [19-23]. However, the observed super-resolutions are limited to a few-photon regime with an extremely low signal-to-noise ratio (SNR), especially in a noisy environment for the entangled photon case [24-26]. On the other hand, actual sensing applications of radars, lidars, gravitation-wave detection [17], Sagnac gyroscopes [27], inertial navigation [28], geodesy [29], magnetometry [30], bio-medical imaging [31], etc. require a high-power light source for high SNR. Thus, PBW (or N00N)-based quantum sensing shows a contradictory power limit to beat its classical counterpart. Moreover, Mach-Zehnder interferometer (MZI)-based quantum sensing technologies with N00N states are suffering from unwanted intensity products based on split photons, resulting in lower fringe visibility [7,32,33]. Here, we experimentally demonstrate classically excited super-resolution using phase-controlled coherent light in a quantum eraser scheme using single photons as well as cw light, where the bedrock of the quantum eraser [34] is a single photon's self-interference [35]. Thus, the observed super-resolution is compatible with conventional sensing technologies based on coherence optics.

The quantum eraser [36-40] has been intensively studied to understand the fundamental physics of the wave-particle duality in quantum mechanics, where the photon's nature is post-determined by measurements. Thus, the key aspect of the quantum eraser is known for the violation of the cause-effect relation [36]. Recently, a coherence solution of the quantum eraser has been analytically derived using the wave nature of a photon without violating quantum mechanics, where the cost to pay for the violation of the cause-effect relation is 50% photon loss by the polarization-basis selection process [34,41]. In that sense, the cause-effect relation is controversial due to the reduced measurement events. The quantum eraser is for the first-order intensity correlation in an interferometric system such as an MZI [37-39,41], even though it has been initially proposed for the entangled photon pairs [36]. Due to the equivalence between quantum and coherent approaches for the first-order intensity correlation [42], the coherence approach is beneficial to understanding the underlying physics of the quantum eraser [37,41]. Thus, the quantum eraser does not have to be limited to a single or entangled-photon pair but instead can be extended to even a continuous-wave (cw) regime [43] due to the intrinsic property of MZI's self-interference [35]. For the present super-resolution, the quantum eraser scheme is modified to be multiple output ports of MZI, where part of them is phase-controlled for the fringe shift. This fringe manipulation of the first-order intensity correlation has already been demonstrated using coherent photons in a non-interferometer scheme for the same super-resolution up to n=30 [19,22]. A general solution of the observed super-resolution has already been found theoretically [44], and the present paper is for the experimental proof of it.

**Result**

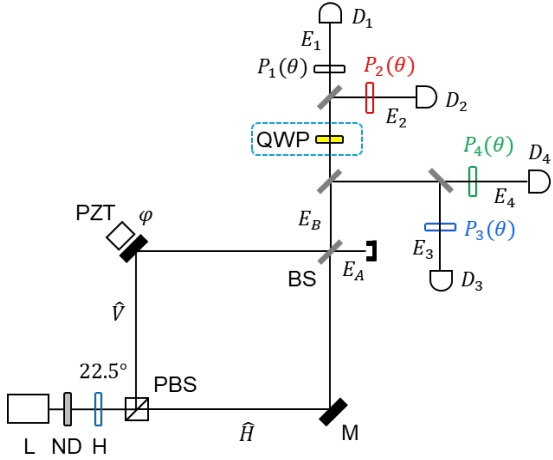

**Fig. 1. Schematic of coherently excited photonic de Broglie waves in a delayed-choice quantum eraser.** L: laser, ND: neutral density filter, H: half-wave plate, PBS: polarizing beam splitter, PZT: piezo-electric transducer, M: mirror, BS: nonpolarizing 50/50 beam splitter, QWP: quarter-wave plate, P: polarizer, D: single photon detector, H: horizon polarization, V: vertical polarization. The rotation angle of P is $\theta = 45°$.

Figure 1 shows the schematic of the coherently excited super-resolution using phase-controlled coherent photons (or cw lights) in a delayed-choice quantum eraser scheme [34,44]. For the phase manipulations of the split output photons, a quarter-wave plate (QWP) is inserted into one of the split output ports of the MZI: For a complete coherence understanding of the quantum eraser, see refs. 37 and 41. The QWP-induced phase shift between orthogonally polarized photons results in the fringe shift of the quantum eraser because the polarizers (Ps) act differently from the polarization bases [34,44].

For random polarization bases of a coherent photon or field, a 22.5°-rotated half-wave plate (HWP) is inserted before entering the quantum eraser composed of a polarizing beam splitter (PBS), 50/50 nonpolarizing beam splitter (BS), and polarizer (P) (see Methods). For the single-photon regime, a set of neutral density (ND) filters is added, resulting in Poisson-distributed coherent photons whose mean photon number $\langle n \rangle$ is far less than unity [34]. By the coincidence detection between output photons, the number (n) of bunched input photons is post-determined, where Fig. 1 is for n=1~4 (see Methods) [34]. In Fig. 1, the QWP-induced fringe shift of the detected photons for the first-order intensity correlation has been theoretically discussed for the super-resolution [44]. Similarly, such a fringe shift of phase-controlled input photons has already been demonstrated for the same super-resolution in a non-interferometric scheme [19,22]. Here, our goal is not for the nonlocal quantum feature but for the PBW-like super-resolution, satisfying the Heisenberg limit overcoming SQL. The super-sensitivity is for a completely separated subject, where the observed PBWs [6-8] have no direct relation with super-sensitivity [11,13].

The polarizer in Fig. 1 plays a critical role in the quantum eraser, where the polarization bases of a single photon enact differently to the first-order intensity correlation [34,39]. Due to the self-interference of a single photon in MZI [35], the input photon number entering the quantum eraser does not affect the fringe, unless a photon-resolving detector is used [9,21,45]. Even in this case, the side effect affected by split photons is inevitable, resulting in degradation of the fringe visibility [19,22]. In our experiments, the single photon counting module (SPCM; Excelitas AQRH-15) cannot resolve photon numbers. Thus, the conventional multiple-intensity product scheme is adopted [8,19,22,33]. Particularly in Fig. 1, only one MZI output port is chosen for the multiple-intensity products. In general, the phase control of the MZI output photons cannot affect fringes due to Born's rule. Very recently, a polarization-dependent phase control of a photon has been studied for the quantum eraser to enable super-resolution [44]. For the phase control of the output photons, a quarter-wave plate (QWP) is inserted before the polarizer to induce $\pm\pi/2$ phase-shifted fringes compared to the non-QWP scheme. Based on this QWP-based phase control of individual quantum erasers in Fig. 1, intensity products with and without QWP show super-resolution (see Fig. 2) [44]. Compared with the classical [19, 22] and quantum [3-14] regimes of super-resolution observed, the present one is robust and compatible with conventional sensing technologies due simply to the MZI-based quantum eraser scheme [37,41,43,44].

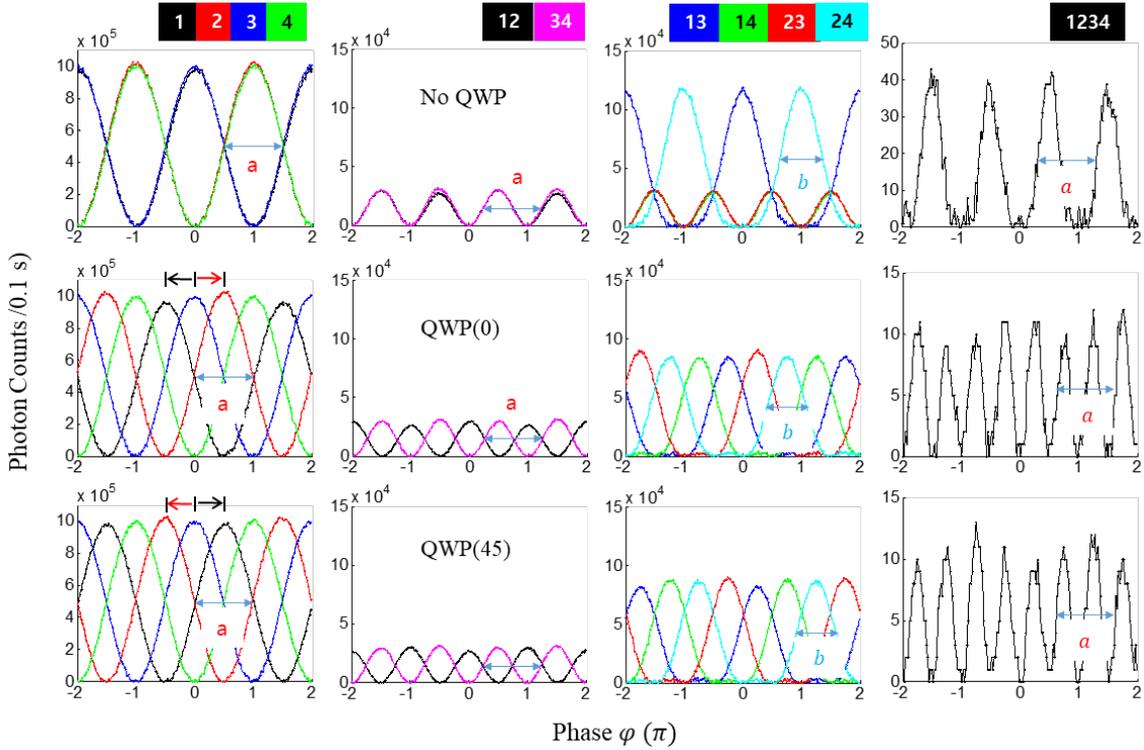

**Fig. 2. Experimental demonstrations of super-resolution using coherent input photons in Fig. 1.** (Top row) without QWP. (Middle row) QWP at $0°$ rotation of the fast axis. (Bottom row) QWP at $45°$ rotation of the fast axis. $a:b = 1:\frac{1}{\sqrt{2}}$ for SQL in classical physics.

Figure 2 shows the experimental results of Fig. 1 for the super-resolution satisfied by PBW-like quantum features [6-11]. The photon counts in Fig. 2 are the measured photon number by SPCMs ($D_{j=1\sim4}$) via a coincidence counting unit (CCU; Altera DE2) for 0.1 s of coincidence measurement time. For the coincidence measurements, a piezo-electric transducer (PZT) is continuously scanned for $-2\pi \leq \varphi \leq 2\pi$ in a forward scanning mode, satisfying a single-shot measurement. Each curve has 360 data points for the $4\pi$ PZT scan range, whose total scanning time is 36 s. For the data collection in each curve, the MZI of Fig. 1 is slightly influenced by air turbulence in a normal lab condition, where the MZI has not been actively or passively controlled. The shaded numbers on top of each panel indicate individual SPCMs for either single or joint measurements.

The top row of Fig. 2 represents the quantum eraser without QWP [34], whereas the other rows are with QWP at different rotation angles. The left-end column of Fig. 2 shows the first-order intensity correlations of individual output photons detected by all SPCMs. The middle two columns are for the second-order intensity correlations jointly measured between any two SPCMs. The right-end column shows the fourth-order intensity correlations among all four SPCMs. All measured data in Fig. 2 satisfy the coincidence detection. As the intensity-product order increases from the left to the right column, the observed photon counts drastically decrease due to the Poisson statistics whose decreasing rate of (n+1) to n bunched photons is $\sim 10^{-1}$. Unlike the conventional quantum approach based on $\chi^{(2)}$ nonlinear optics suffering from phase mismatching due to the propagation walk-off in the birefringent medium and misalignment in a photon collection scheme, the errors in Fig. 2 are mostly from air turbulence of MZI. All data in Fig. 2 are single-shot measurements during the PZT scan.

As a reference of the quantum eraser without QWP in the top row of Fig. 2, related fringe resolutions are denoted by 'a' and 'b,' where 'a' is the diffraction limit of classical physics in the unit of full-width-at-half maximum (FWHM). The scale factor $1/\sqrt{2}$ of 'b' to 'a' satisfies SQL for n=2. The middle-left panel shows the 2nd–order intensity correlation between SPCMs, where the scale factor 1/2 is beyond the SQL. This is, of course, a usual resolution enhancement between MZI output ports resulting from the out-of-phase relation. The 4th–order intensity correlation in the right-end panel also shows a nonclassical feature overcoming SQL, whose scale factor is less than $1/\sqrt{4}$ for n=4. Instead, the resolution enhancement of this 4th–order intensity product follows SQL, while the 2nd–order intensity product of the middle-left panel shows the scale factor of $1/\sqrt{2}$ (see Fig. 2). As a result, the fringe quadrupling between n=4 and n=1 is not satisfied, resulting in no super-resolution. The opposite fringe patterns between SPCMs '1 (black)' and '2 (red)' as well as '3 (blue)' and '4 (green)' in the left-end panel are due to the BS-caused sign reversal in $\hat{H}$ polarization basis in the quantum eraser scheme [34,41], (see Analysis). This fringe relation is the same as the usual MZI case, even though MZI physics lies in the BS matrix-caused $\pi/2$-phase shift between reflected and transmitted photons [46].

The middle row with QWP whose fast-axis is vertical (FA-V), i.e., slow-axis horizontal (SA-H), demonstrates the PBW-like super-resolution with n-proportional fringe multiplication. Thus, the measured phase resolutions of the first (n=1), second (n=2), and fourth (n=4) intensity correlations are linearly enhanced as the intensity-product order *n* increases, as in PBWs [8]. The fringe quadrupling in the right-end panel is the witness of the super-resolution as observed by using phase-controlled coherent photons [19,22] or higher-order entangled-photon pairs of N00N states [4,9,11,12]. Compared to the top row, the first-order intensity products in the middle row show $\pm\pi/2$ fringe shifts by the action of QWP. This is the unprecedented result, where the first-order intensity fringes of '1' and '2' are $\mp\pi/2$ phase-shifted compared to '3' and '4,' respectively (see Analysis). Thus, the physical origin of the fringe doubling (quadrupling) phenomena in the second-order (fourth-order) intensity correlations is the QWP-induced phase gain to the vertical polarization basis of the input photon [44]. This kind of quantum eraser-based super-resolution is unprecedented in both coherence and quantum optics, where the experimental results perfectly match the theory [44]. Because the Heisenberg limit is the definite quantum feature, the middle row in Fig. 2 is the witness of the classically excited quantum features of super-resolution via phase manipulations of a coherent photon in the output port of MZI.

The bottom row is for a 45°-rotated QWP to the counterclockwise direction. Unlike the SA-H QWP (0) in the middle row, the first-order intensity-fringe shifts of '1' and '2' are swapped (or reversed) with each other, as shown in the left-end panel. However, these opposite fringe shifts between '1' and '2' do not affect the second-order intensity products due to the same out-of-phase relation between them, as shown in the middle-left panel. Thus, the fourth-order intensity correlation in the right-end panel shows the same fringe doubling of the second-order intensity correlation in the middle-left panel, satisfying the Heisenberg limit in phase resolution. As observed, the physical reason for fringe doubling in the second-order intensity correlation in the middle-left panel is due to the out-of-phase between the first-order intensity fringes in the left-end panel. Likewise, the fourth-order intensity fringe doubling in the right-end panel is due to the out-of-phase between the second-order intensity fringes in the middle-left panel, too. For a complete understanding of the bottom row, refer to the generalized coherence solution with an arbitrary angle of QWP (see also Analysis) [44].

The QWP at the FA-V induces the right circularly polarized output photon for the orthogonally polarized input photons [1]. For the QWP at the FA-H, the output photons are left circularly polarized [1]. The middle row is for the FA-V (i.e., SA-H) QWP, resulting in a $-\pi/2$ phase shift to the vertical component of a single photon (see Eqs. (1) and (2) in Analysis). For the last BS-reflected photon toward detectors '3' and '4,' the $\hat{H}$ component experiences a $\pi$ phase shift by the mirror image, resulting in the sign change. As analyzed in *Analysis*, thus, the four output photons have a sine and cosine relation, resulting in the unprecedented equally phase-shifted fringes in the first-order intensity correlations of the quantum eraser. These equally spaced fringes have also been observed using phase-controlled input photons in a non-interferometric scheme for the demonstration of super-resolution, satisfying HL [19,22].

*Analysis*
For the wave nature of quantum mechanics, the output photons from the MZI in Fig. 1 are represented as $E_A = \frac{iE_0}{2}(\hat{H} + \hat{V}e^{i\varphi})$ and $E_B = \frac{E_0}{2}(\hat{H} - \hat{V}e^{i\varphi})$ by the BS matrix representations [46], where $E_0$ is the amplitude of a single photon. $\hat{H}$ and $\hat{V}$ are unit vectors of horizontal and vertical polarizations, respectively. The final photon amplitudes modified by the polarizer rotated at $\theta$ from the horizontal axis are represented as $E_1 = \frac{E_0}{4}(\hat{H}cos\theta - \hat{V}sin\theta e^{i\varphi})\hat{p}$ and $E_2 = \frac{-iE_0}{4}(\hat{H}cos\theta + \hat{V}sin\theta e^{i\varphi})\hat{p}$ for the detectors '1' and '2.' $\hat{p}$ is the axis of the polarizers from the horizontal axis to the counterclockwise direction. For $E_2$, a $\pi$-phase shift for $\hat{H}$ only is considered for the mirror image. Here, the meaningless $\hat{H}$ and $\hat{V}$ are just to indicate the photon's origin. Thus, the corresponding intensities of the quantum erasers become $\langle I_1 \rangle = \frac{I_0}{16}\langle 1 - sin2\theta cos\varphi \rangle$ and $\langle I_2 \rangle = \frac{I_0}{16}\langle 1 + sin2\theta cos\varphi \rangle$, where the global phase has no effect in the intensity due to the Born's rule that the measurement is the absolute square of the probability amplitude.

Similarly, the amplitudes for detectors '3' and '4' are represented as follows:

$$E_3 = \frac{-E_0}{4}(\hat{H}cos\theta - \hat{V}sin\theta e^{i\varphi})\hat{p}, \qquad (1)$$

$$E_4 = \frac{-iE_0}{4}(\hat{H}cos\theta + \hat{V}sin\theta e^{i\varphi})\hat{p}. \qquad (2)$$

Corresponding intensities are $\langle I_3 \rangle = \frac{I_0}{16}\langle 1 - sin2\theta cos\varphi \rangle$ and $\langle I_4 \rangle = \frac{I_0}{16}\langle 1 + sin2\theta cos\varphi \rangle$. As observed in the left panel of the top row in Fig. 2, thus, the first-order intensity products are analytically confirmed for $\theta = \pi/4$ without QWP: $\langle I_1 \rangle = \langle I_3 \rangle = \frac{I_0}{16}\langle 1 - cos\varphi \rangle$ and $\langle I_2 \rangle = \langle I_4 \rangle = \frac{I_0}{16}\langle 1 + cos\varphi \rangle$. As a result, the second-order intensity products are $\langle C_{12}^{(2)} \rangle = \langle C_{34}^{(2)} \rangle = \frac{I_0^2}{64}\langle sin^2\varphi \rangle$, where $\langle C_{ij}^{(2)} \rangle = \langle I_i I_j \rangle$, as observed in the middle-left panel of the top row. Here, each amplitude of the two-photon case is multiplied by a factor of $\sqrt{2}$. Similarly, $\langle C_{13}^{(2)} \rangle =$ $\frac{I_0^2}{64}\langle(1-cos\varphi)^2\rangle$, $\langle C_{24}^{(2)}\rangle = \frac{I_0^2}{64}\langle(1+cos\varphi)^2\rangle$, and $\langle C_{23}^{(2)}\rangle = \langle C_{14}^{(2)}\rangle = \langle C_{12}^{(2)}\rangle$ are obtained, as observed in the middle-right panel of the top row in Fig. 2. Thus, the SQL is satisfied for $\langle C_{13}^{(2)}\rangle$ and $\langle C_{24}^{(2)}\rangle$, resulting in $\sqrt{2}$ enhancement in resolution. The fourth-order intensity correlation in the right panel is, however, from the middle-left panel, resulting in $\langle C_{1234}^{(4)}\rangle = \langle C_{12}^{(2)}C_{34}^{(2)}\rangle = \langle C_{13}^{(2)}C_{24}^{(2)}\rangle = \frac{I_0^4}{256}\langle sin^4\varphi\rangle$. Thus, the out-of-phase relation between the first-order intensity correlations in the right-end panel of the top row in Fig. 2 is not for SQL, resulting in $\langle C_{12}^{(2)}\rangle \ne \langle C_{13}^{(2)}\rangle$ and $\langle C_{1234}^{(4)}\rangle \ne \langle I_2^4\rangle$.

By the QWP whose rotation angle is 0° for the SA-H, the amplitudes of $\boldsymbol{E}_1$ and $\boldsymbol{E}_2$ are modified as follows:

$$\boldsymbol{E}_{1Q} = \frac{E_0}{4}(\widehat{H}cos\theta - i\widehat{V}sin\theta e^{i\varphi})\hat{p}, \tag{3}$$

$$\boldsymbol{E}_{2Q} = \frac{-iE_0}{4}(\widehat{H}cos\theta + i\widehat{V}sin\theta e^{i\varphi})\hat{p}, \tag{4}$$

where SA-H QWP induces $\pi/2$ phase gain to the $\widehat{V}$ component. Thus, the corresponding intensities measured by SPCMs are as follows:

$$\langle I_{1Q}\rangle = \frac{I_0}{16}\langle 1 + sin2\theta sin\varphi\rangle, \tag{5}$$

$$\langle I_{2Q}\rangle = \frac{I_0}{16}\langle 1 - sin2\theta sin\varphi\rangle, \tag{6}$$

For diagonally rotated polarizers at $\theta = \pi/4$, $\langle I_{1Q}\rangle = \frac{I_0}{16}\langle 1 + sin\varphi\rangle$, $\langle I_{2Q}\rangle = \frac{I_0}{16}\langle 1 - sin\varphi\rangle$, $\langle I_3\rangle = \frac{I_0}{16}\langle 1 - cos\varphi\rangle$, and $\langle I_4\rangle = \frac{I_0}{16}\langle 1 + cos\varphi\rangle$ are obtained for the SA-H QWP. Here, the PZT-controlled $\varphi$ is a definite parameter, enabling the single-shot measurement of the quantum eraser. This understanding is quite important to follow the unitary transformation of the BS matrix in the quantum eraser, even though a single BS results in randomness. As a result, $\langle I_{1Q}\rangle$ and $\langle I_{2Q}\rangle$ are $\pm\pi/2$-phase shifted from $\langle I_1\rangle$, respectively. Thus, the fringes observed in the left-end panels of the top and middle rows are analytically confirmed.

The second-order intensity correlation between Eqs. (5) and (6) for the SPCMs '1' and '2' is as follows (see the middle-left panel of the middle row in Fig. 2):

$$\langle C_{1Q2Q}^{(2)}\rangle = \langle I_{1Q}I_{2Q}\rangle = \frac{I_0^2}{64}\langle cos^2\varphi\rangle = \frac{I_0^2}{(2\cdot 64)}\langle 1 + cos2\varphi\rangle, \tag{7}$$

where the amplitude is multiplied by a factor of $\sqrt{2}$ to compensate for the two-photon coincidence case. Likewise, the two-photon intensity product between '3' and '4' is:

$$\langle C_{34}^{(2)}\rangle = \langle I_3 I_4\rangle = \frac{I_0^2}{64}\langle sin^2\varphi\rangle = \frac{I_0^2}{(2\cdot 64)}\langle 1 - cos2\varphi\rangle. \tag{8}$$

Equations (7) and (8) are shown in the middle-left panel of the middle row in Fig. 2. However, the second-order intensity correlation $\langle C_{12}^{(2)}\rangle = \frac{I_0^2}{64}\langle sin^2\varphi\rangle$ between detectors '1' and '2' is the same as Eq. (8), as shown in the middle-left panel of the top row. Thus, the out-of-phase relation is obtained for Eqs. (7) and (8), as observed.

For the fourth-order intensity correlation between Eqs. (7) and (8), the quadrupled fringes are resulted (see the right panel of the middle row in Fig. 2):

$$\langle C^{(4)}_{1Q2Q34}\rangle = \frac{I_0^4}{256}\langle sin^2\varphi cos^2\varphi\rangle = \frac{I_0^4}{4\cdot 256}\langle sin^2(2\varphi)\rangle = \frac{I_0^4}{(8\cdot 256)}\langle 1-cos4\varphi\rangle, \tag{9}$$

where a factor 2 is multiplied to the amplitudes to compensate for the four-photon coincidence case. In Eq. (9), fringe quadrupling is obtained for n=4. The same result is also obtained from the middle-right panel of Fig. 2 for n=2. Thus, the fringe doubling (quadrupling) of the second-order (fourth-order) intensity correlations in the middle row of Fig. 2 are analytically confirmed for the super-resolution, satisfying the Heisenberg limit. The origin of the PBW-like super-resolution in Eq. (9) is the $\pi/2$ phase shift between product bases.

For a generalized SA-H QWP with an arbitrary rotation angle $\xi$, Eqs. (3) and (4) are rewritten as:

$$\mathbf{E}^\xi_{1Q} = \frac{E_0}{2}e^{-i2\xi}(\hat{H}cos\theta - i\hat{V}sin\theta e^{i(\varphi+4\xi)})\hat{p}, \tag{10}$$

$$\mathbf{E}^\xi_{1Q} = \frac{-iE_0}{2}e^{-i2\xi}(\hat{H}cos\theta + i\hat{V}sin\theta e^{i(\varphi+4\xi)})\hat{p}. \tag{11}$$

The corresponding intensities are as follows:

$$\langle I^\xi_{1Q}\rangle = \frac{I_0}{16}\langle 1+sin2\theta sin(\varphi+4\xi)\rangle, \tag{12}$$

$$\langle I^\xi_{2Q}\rangle = \frac{I_0}{16}\langle 1-sin2\theta sin(\varphi+4\xi)\rangle. \tag{13}$$

For $\xi=\theta=45°$, $\langle I^{\xi=45°}_{1Q}\rangle = \frac{I_0}{16}\langle 1-sin\varphi\rangle$ and $\langle I^{\xi=45°}_{2Q}\rangle = \frac{I_0}{16}\langle 1+sin\varphi\rangle$ are resulted. Thus, the fringe swapping between $\xi=0°$ and $\xi=45°$ for '1' and '2' are analytically confirmed for the left-end panels of the middle and bottom rows in Fig. 2. The second-order intensity correlation between Eqs. (12) and (13) is the same as Eq. (7), as shown in the middle-left panel of the bottom row in Fig. 2. For the unchanged $\langle I_3\rangle$ and $\langle I_4\rangle$, thus, the fourth-order intensity correlation is also the same as Eq. (9), as shown in the right-end panel of the bottom row in Fig. 2. As a result, all observed features are analytically confirmed. Thus, the PBW-like quantum features observed in Fig. 2 are understood as coherently achievable definite features using phase manipulations of the quantum eraser. Unlike the nonlocal quantum feature, the observed PBW-like super-resolution for n=1,2,4 can be interpreted as a result of the quantized product bases in the higher-order intensity correlations. For this product-basis quantization, the coherence approach for the wave nature of a photon successfully confirmed the observed quantum features via coherence manipulations of Poisson-distributed photons using linear optics, QWP.

Figure 3 shows cw light-based super-resolution for Fig. 1. For the measurements in Fig. 3, the SPCMs used in Fig. 2 are replaced by photodiodes (Thorlabs, APD-110A), where the intensity products are conducted on a fast (500 MHz) digital oscilloscope (Yokogawa, DL9040) without CCU. The input laser power of 300 μW before entering MZI is adjusted by ND filters. Due to the coherence feature of the quantum eraser as analyzed in *Analysis*, the same super-resolutions as in Fig. 2 are observed. Here, the fringe visibility is caused by the MZI air turbulence in normal lab conditions. In our case, the MZI stability continues for several minutes. The measured intensity ratios of $\langle C^{(2)}_{ij}\rangle$ and $\langle C^{(4)}_{1Q2Q34}\rangle$ to $\langle I_0\rangle$ are satisfied for Eqs. (7)-(9), as expected. Thus, the coherence understanding of the observed cw super-resolution is now completed for the phase-controlled quantum erasers in Fig. 1. Unlike previous nonclassical light-based super-resolution limited to noisy environments, the present method of coherence manipulations of classical light in a quantum eraser scheme demonstrates both new physics and a breakthrough in quantum sensing applications compatible with conventional sensor technologies based on high SNRs ranging from coherent radars to bio-imaging [31]. Due to the phase-controlled intensity products between quantum erasers, the final goal of quantum sensing overcoming SQL has been achieved for super-resolution.

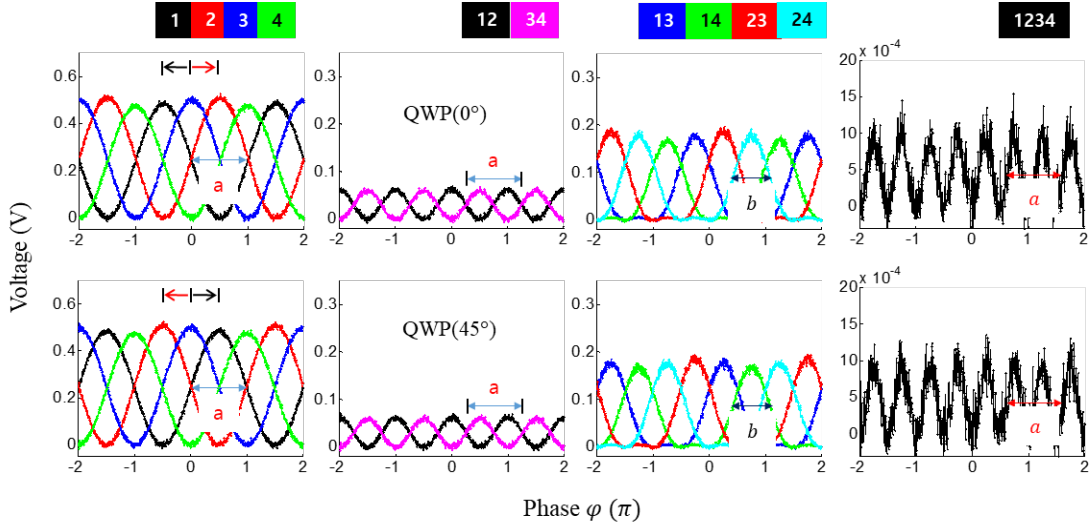

**Figure 3. Experimental demonstrations of super-resolution using cw light for Fig. 1.** (Top row) QWP at 0° rotation of the fast axis. (Bottom row) QWP at 45° rotation of the fast axis. The light's input power to MZI is 300 μW.

**Conclusion**

Using a quarter-wave plate, phase manipulations of coherent photons were conducted in a quantum eraser scheme for the super-resolution overcoming SQL. The PBW-like quantum features of super-resolutions were observed for the higher-order intensity correlations between phase-controlled and uncontrolled quantum erasers, whose intensity product fringes were intensity-order proportional, satisfying the Heisenberg limit of quantum sensing in phase resolution. In addition, the same super-resolution was observed in a cw regime, where the quantum eraser has no distinction between a single photon and cw light due to the intrinsic MZI property of the single photon's self-interference. Corresponding analyses confirmed the observed data, where the origin of the super-resolution was in the QWP-induced fringe shifts of the quantum eraser, resulting in the equally shifted fringes of the first-order intensity correlations. Compared to the practical limitations of PBW (or N00N state)-based quantum sensing in both ultralow generation efficiency and noisy environments, the present method showed potential due to its compatibility with coherence optic-based sensing technologies.

**References**

1. Pedrotti, F. L., Introduction to Optics. 3rd Ed. (Pearson Education, Inc. 2007).
2. Braginskii, V. G. & Vorontsov, Yu. I. Quantum-mechanical limitations in macroscopic experiments and modern experimental technique. Sov. Phys. Usp. **17**, 644-650 (1975).
3. Giovannetti, V., Lloyd, s. & Maccone, L. Advances in quantum metrology. Nat. Photon. **5**, 222-229 (2011).
4. Dowling, J. P. Quantum optical metrology-the lowdown on high-N00N states. Contemp. Phys. **49**, 125-143 (2008).
5. Kuzmich, A. & Mandel, L. Sub-shot-noise interferometric measurements with two-photon states. Quantum Semiclass. Opt. **10**, 493-500 (1998).
6. Jacobson, J., Gjörk, G., Chung, I. & Yamamato, Y. Photonic de Broglie waves. Phys. Rev. Lett. **74**, 4835–4838 (1995).
7. Edamatsu, K., Shimizu, R. & Itoh, T. Measurement of the photonic de Broglie wavelength of entangled photon pairs generated by parametric down-conversion. Phys. Rev. Lett. **89**, 213601 (2002).
8. Walther, P. et al. Broglie wavelength of a non-local four-photon state. Nature **429**, 158–161 (2004).

9. Giovannetti, V., Lloyd, S. & Maccone, L. Quantum-enhanced measurements: beating the standard quantum limit. Science **306**, 1330–1336 (2004).
10. Leibfried, D. et al. Toward Heisenberg-limited spectroscopy with multiparticle entangled states. Science **304**, 1476–1478 (2004).
11. Nagata, T., Okamoto, R., O'Biran J. L., Sasaki, K. & Takeuchi, S. Beating the standard quantum limit with four-entangled photons. Science **316**, 726-729 (2007).
12. Zhang, J. et al. N00N states of nine quantized vibrations in two radial modes of a trapped ion. Phys. Rev. Lett. **121**, 160502 (2018).
13. Daryanoosh, S., Slussarenko, S., Berry, D. W., Wiseman, H. M. & Pryde, G. J. Experimental optical phase measurement approaching the exact Heisenberg limit. Nature Communi. **9**, 4606 (2018).
14. Berni, A. A., Gehring T., Nielsen, Bo. M., Händchen, V., Paris, M. G. A. & Andersen, U. L. Ab initio quantum-enhanced optical phase estimation using real-time feedback control. Nature Photon. **9**, 577-581 (2015).
15. Schnabel, R. Squeezed states of light and their applications in laser interferometers. Phys. Rep. **684**, 1-51 (2017).
16. Lawrie, B. J., Lett. P. D., Marino, A. M. & Pooser, R. C. Quantum sensing with squeezed light. ACS Photon. **6**, 1307-1318 (2019).
17. Tse, M. et al. Quantum-enhanced advanced LIGO detectors in the era of gravitational-wave astronomy. Phys. Rev. Lett. **123**, 231107 (2019).
18. Cunubum V, et al. Experimental metrology beyond the standard quantum limit for a wide resource range. npj Quant. Info. **9**, 20 (2023).
19. Resch. K. J., Pregnell, K. L., Prevedel, R., Gilchrist, A., Pryde, G. J., O'Brien, J. L. & White, A. G. Time-reversed and super-resolving phase measurements. Phys. Rev. Lett. **98**, 223601 (2007).
20. Cohen, L., Istrati, D., Dovrat, L. & Eisenberg, H. S. Super-resolved phase measurements at the shot noise limit by parity measurement. Opt. Exp. **22**, 1194511953 (2014).
21. Gao, Y., Anisimov, P. M., Wildfeuer, C. F., Luine, J., Lee, H. & Dowling, J. P. Super-resolution at the shot-noise limit with coherent states and photon-number-resolving detectors. J. Opt. Soc. Am. **27**, A170-A174 (2010).
22. Kothe, C., Björk, G. & Bourennane, M. Arbitrarily high super-resolving phase measurements at telecommunication wavelengths. Phys. Rev. A **81**, 063836 (2010).
23. Gerry, C. C. & Mimih, J. Heisenberg-limited interferometry with pair coherent states and parity measurements. Phys. Rev. A **82**, 013831 (2010).
24. Lopaeva, E. D., Berchera, I. R., Degiovanni, I. P., Olivares, S., Brida, G. & Genovese, M. Experimental realization of quantum illumination. Phys. Rev. Lett. **110**, 123603 (2013).
25. Lloyd, S. Enhanced sensitivity of photodetection via quantum illumination. Science **321**, 1463–1465 (2008).
26. Gregory, T., Moreau, P. A., Toniinelli, E. & Padgett, M. J. Imaging through noise with quantum illumination. Sci. Adv. **6**, eaay2652 (2020).
27. Hurst, R.B.; Stedman, G.E.; Schreiber, K.U.; Thirkettle, R.J.; Graham, R.D.; Rabeendran, N.; Wells, J.-P.R. Experiments with a 834 m2 ring laser interferometer. J. Appl. Phys. **105**, 113115 (2009).
28. Chow, W.W.; Gea-Banacloche, J.; Pedrotti, L.M.; Sanders, V.E.; Schleich, W.; Scully, M.O. The ring laser gyro. Rev. Mod. Phys. **57**, 61–104 (1985).
29. Schreiber, K.U.; Klügel, T.; Wells, J.-P.R.; Hurst, R.B.; Gebauer, A. How to detect the Chandler and the annual wobble of the Earth with a large ring laser gyroscope. Phys. Rev. Lett. **107**, 173904 (2011).
30. Budker, D. & Romalis, M. Optical magnetometry. Nature Phys. **3**, 227-234 (2007).
31. Fujimoto, J. G., Pitris, C., Boppart, S. A. & Brezinski, M. E. Optical coherence tomography: an emerging technology for biomedical imaging and optical biopsy. Neoplasia **2**, 9-25 (2000).
32. Xiang, G. Y., Higgins, B. L., Berry, D. W., Wiseman, H. M. & Pryde, G. J. Entanglement-enhanced measurement of a completely unknown optical phase. Nature Photon. **5**, 43-47 (2011).

33. Sun, F. W., Liu, B. H., Gong, Y. X., Huang, Y. F., Ou, Z. Y. & Guo, G. C. Experimental demonstration of phase measurement precision beating standard quantum limit by projection measurement. EPL **82**, 24001 (2008).
34. Kim, S. & Ham, B. S. Observations of the delayed-choice quantum eraser using coherent photons. Sci. Rep. **13**, 9758 (2023).
35. Grangier, P., Roger, G. & Aspect, A. Experimental evidence for a photon anticorrelation efect on a beam splitter: A new light on single-photon interferences. Europhys. Lett. **1**, 173–179 (1986).
36. Scully, M. O. & Drühl, K. Quantum eraser: A proposed photon correlation experiment concerning observation and "delayed choice" in quantum mechanics. Phys. Rev. A **25**, 2208–2213 (1982).
37. Kim, Y.-H., Yu, R., Kulik, S. P. & Shih, Y. Delayed, "choice" quantum eraser. Phys. Rev. Lett. **84**, 1–5 (2000).
38. Herzog, T. J., Kwiat, P. G., Weinfurter, H. & Zeilinger, A. Complementarity and the quantum eraser. Phys. Rev. Lett. **75**, 3034–3037 (1995).
39. Jacques, V. et al. Experimental realization of wheeler's delayed-choice Gedanken experiment. Science **315**, 966–978 (2007).
40. Ma, X.-S., Kofer, J. & Zeilinger, A. Delayed-choice gedanken experiments and their realizations". Rev. Mod. Phys. **88**, 015005 (2016).
41. Ham, B. S. Coherence interpretation of nonlocal quantum correlation in a delayed-choice quantum eraser. Arxiv:2206.05358v3 (2022).
42. Stöhr, J. Overcoming the diffraction limit by multi-photon interference: a tutorial. Adv. Opt. & Photon. **11**, 215-313 (2019).
43. Ham, B. S. Observations of the delayed-choice quantum eraser in a macroscopic system. arXiv:2206.05358v3 (2022).
44. Ham, B. S. Phase-controlled coherent photons for the quantum correlations in a delayed-choice quantum eraser scheme. arXiv:2310.13217 (2023).
45. Cheng, R., Zhou, Y., Wang, S., Shen, M., Taher, T. & Tang, H. X. A 100-pixel photon-number-resolving detector unveiling photon statistics. Nature Photon. **17**, 112-119 (2023).
46. Degiorgio, V. Phase shif between the transmitted and the refected optical felds of a semirefecting lossless mirror is $\pi/2$. Am. J. Phys. **48**, 81–82 (1980).

**Methods**

The noninterfering Mach-Zehnder interferometer (MZI) in Fig. 1 comprises PBS and BS. By the PBS, the randomly polarized input photons are deterministically split into the MZI output ports. By the 50/50 unpolarizing BS of the MZI, thus, both MZI output ports provide random polarization bases of a photon, resulting in no fringes. For the delayed-choice quantum eraser, a polarizer is inserted in each MZI output port. For the PBW-like quantum feature of super-resolution, one MZI output port is divided into four different quantum erasers. Two of them are phase-controlled with a QWP. Finally, four detectors of the four quantum erasers are connected to a coincidence counting unit (CCU; Altera, DE2) for higher-order intensity correlation measurements via coincidence measurements. By CCU, a particular input photon number (n=1~4) is post-determined from Poisson-distributed photons. Due to the Poisson statistics, however, a few percent error is inevitable for coincidence detection because our SPCMs cannot resolve photon numbers. For the cw experiments in Fig. 3, each data point is 30-sample averaged by the internal function of the oscilloscope (Yokogawa, DL9040), where the input light power before entering MZI is 300 µW to avoid detector (Thorlabs, APD-110A) saturation.

**Funding:** This research was supported by the MSIT (Ministry of Science and ICT), Korea, under the ITRC (Information Technology Research Center) support program (IITP 2023-2021-0-01810) supervised by the IITP (Institute for Information & Communications Technology Planning & Evaluation). BSH also acknowledges that this work was also supported by GIST GRI-2023.

**Author contribution:** SK conducted experiments. BSH supervised the experiments, analyzed the data, prepared the theory, and wrote the paper.

**Competing Interests:** The author declares no competing interest.
**Data Availability Statement:** All data generated or analyzed during this study are included in this published article.